\documentclass{llncs}
\usepackage{graphicx,amssymb,amsmath}
\usepackage{tikz}
\usepackage[disable]{todonotes}
\begin{document}

\newif\ifarxiv 
\arxivtrue
\graphicspath{{figures/}}

\let\doendproof\endproof
\renewcommand\endproof{~\hfill\qed\doendproof} 
 
\mainmatter

\title{On Upward Drawings of Trees on a Given Grid\thanks{Work of the authors supported in part by NSERC.}}

\author{
Therese Biedl$^\S$
\and
Debajyoti Mondal$^\dagger$
}

\institute{  
$^\S$Cheriton School of Computer Science,  University of Waterloo, Canada \\
$^\dagger$Department of Computer Science, University of Saskatchewan, Canada \\
\email{biedl@uwaterloo.ca, dmondal@cs.usask.ca}
}


\maketitle

\begin{abstract} 
 Computing a minimum-area planar straight-line drawing of a graph is known to be NP-hard for planar graphs,
 even when restricted to outerplanar graphs. However, the complexity question is open for trees. Only a few
 hardness results are known for straight-line drawings of trees under various restrictions such as edge length 
 or slope constraints. On the other hand, there exist polynomial-time algorithms for computing  minimum-width
 (resp., minimum-height) upward drawings of trees, where the height (resp., width) is unbounded.  
  
 In this paper we take a major step in understanding the complexity of the area minimization problem for  
 strictly-upward drawings of trees, which is one of the most common styles for drawing rooted trees. 
 We prove that given a rooted tree $T$ and a $W\times H$ grid, it is NP-hard to 
 decide whether $T$ admits a strictly-upward (unordered) drawing in the given grid.
 The hardness result holds both in polyline and straight-line drawing settings. 
\end{abstract}

\section{Introduction} 

Drawing planar graphs on a small integer grid is an active research area~\cite{handbook},
 which is motivated by various practical needs such as for VLSI circuit layout and
 small-screen visualization.  Trees are  one of the most studied graph classes in this context.  
 While computing a minimum-area planar straight-line drawing
 of an arbitrary planar graph is known to be NP-complete~\cite{KW}, 
 even for planar graphs with bounded pathwidth~\cite{KW} and outerplanar graphs
\cite{DBLP:conf/icalp/Biedl14}, the problem seems
 very intriguing for trees.  
 
In this paper we examine \emph{rooted and unordered trees}, i.e., one of the vertices is designated
 as the root and the left to right order of the children can be chosen arbitrarily.  
 A natural way to display such a tree is to compute a \emph{(strictly) upward} drawing, where 
 each vertex is mapped to an integer grid point such that the parents have (strictly)
 larger $y$-coordinates than their children, each edge is drawn with $y$-monotone 
 polylines with bends at integer grid points, and no two edges cross except possibly
 at their common endpoint. The \emph{width, height} and \emph{area}
 of the drawing are respectively the width, height, and area  
 of the smallest axis-parallel rectangle that encloses the drawing.  
 In a \emph{straight-line (strictly) upward drawing}, the edges are restricted to be straight line segments.

We refer the reader to~\cite[Chapter 5]{handbook} or \cite{DF14} for a survey on small-area drawings of trees.
 Here we review only the results that focus on exact minimization. 
 In the fixed-embedding setting, there exist polynomial-time algorithms to compute minimum-area drawings for 
 certain classes of planar graphs~\cite{DBLP:conf/icalp/Biedl14,DBLP:journals/jgaa/MondalNRA11},
namely, those that simultaneously have bounded treewidth and bounded face-degrees.
 On the other hand, the problem becomes NP-hard as soon as  one of the above constraints is dropped~\cite{DBLP:conf/icalp/Biedl14}.  The intractability of minimum-area tree drawings has  been well established 
 under some edge length and slope restrictions, e.g., when edges 
 must be drawn with unit length or the slopes of the edges 
 in the drawing must belong to one of the $(k/2)$ slopes determined by a $k$-grid.  
 Note that an orthogonal grid is a $4$-grid.  
 Determining whether a tree can be drawn on a given $k$-grid, where $k\in \{4,6,8\}$
 with edges of unit length is an NP-complete problem~\cite{DBLP:journals/ipl/BhattC87,DBLP:journals/ipl/Gregori89,DBLP:journals/jgaa/BachmaierM13}. 
 Similar hardness results hold also for ordered trees under various aesthetic requirements~\cite{DBLP:conf/gd/BrunnerM10}.

Not much is known about minimum-area upward drawings of trees.
 Trevisan~\cite{DBLP:journals/ipl/Trevisan96} showed that 
 every complete tree (resp., Fibonacci tree) with $n$ vertices admit a 
 strictly-upward straight-line drawing in 
 $n+O(\log n\sqrt n)$ (resp., $1.17n ++O(\log n\sqrt n)$) area,
 and  conjectured that exact minimization may be possible
 in polynomial time. 
 Trevisan mentioned that the problem of minimum-area strictly-upward
 drawing of a complete tree is a `sparse problem'.
Therefore, proving this NP-hard would imply $P=NP$
 by Mahaney's theorem~\cite{DBLP:journals/jcss/Mahaney82}.
\todo{TB: I've shortened this a bit to save space.}

Interestingly, there exist polynomial-time algorithms for
 computing  minimum-width upward drawings of trees, where the height 
 is unbounded~\cite{DBLP:journals/corr/Biedl15}, and minimum-height 
 drawings where
 the width is unbounded~\cite{DBLP:journals/jgaa/AlamSRR10} (even
 for non-upward drawings~\cite{DBLP:conf/walcom/MondalAR11}).  
 Marriott and Stuckey~\cite{DBLP:journals/jgaa/MarriottS04} showed that minimizing the width of 
 a strictly-upward straight-line drawing is NP-hard under the additional constraint that 
 the $x$-coordinate of a parent is the average of the $x$-coordinates of its
 children. This additional requirement implies that   a vertex with 
 a single child must be placed directly above the child.
 Minimizing the width is known to be NP-hard even for ordered tree
 drawings under several other constraints~\cite{DBLP:journals/acta/SupowitR82}. 

We show that computing a strictly-upward drawing of a tree that resides  
 in a given $W\times H$-grid is NP-hard. Formally, we consider the following problem:
\begin{enumerate}
\itemsep -1pt
\item[]\textbf{Problem:} \textsc{Strictly-Upward Tree Drawing} (SUTD)
\item[]\textbf{Instance:} An unordered rooted tree $T$, and two natural numbers $W$ and $H$. 
\item[]\textbf{Question:} Does $T$ admit a strictly-upward (polyline or straight-line) drawing with width at most $W$ and height at most $H$?
\end{enumerate}
 
\section{NP-hardness}  
 
We prove the NP-hardness of SUTD by a reduction from the following problem:
\begin{enumerate}
\itemsep -1pt
\item[]\textbf{Problem:} \textsc{Numerical 3-Dimensional Matching} (N3DM)
\item[]\textbf{Instance:} Positive integers $r_i,g_i,b_i$, where $1\le i\le k$, and an integer $B$ such that 
  $\sum_{i=1}^k (r_i+b_i+g_i) = k\cdot B$.  
\item[]\textbf{Question:} Do there exist permutations $\pi$ and $\pi'$  of $\{1,\dots,k\}$ such that
  $r_{\pi(i)}+b_i+g_{\pi'(i)} = B$ for all $1\leq i\leq k$?
\end{enumerate}
N3DM is strongly NP-complete~\cite{gANDj}, 
and remains NP-complete even if we require all
 $b_i$'s to be odd, and the $b_i$'s are large and the $g_i$'s are huge
relative to the $r_i$'s.  (More precisely, $3k^c \le r_i \le 4k^c$, $k^{2c} \le b_i \le k^{2c}+k^c$, 
 and $k^{4c} \le g_i \le k^{4c}+k^c$, where $c>1$ is a constant.) 
 See 
\ifarxiv
the appendix
\else
the full version~\cite{arxiv} 
\fi
for further details.

\paragraph{\bf Idea of the Reduction:} 
One crucial ingredient is to construct a tree whose height matches the 
height-bound $H$ of the drawing; this determines the layer for all vertices 
on paths 
 to the deepest leaves because in a strictly-upward 
drawing every edge must lead to a layer farther down.
Another crucial ingredient is to add so
many vertices to the tree  that (other than on the topmost layer) all
grid-points must have a vertex on them.  The next ingredient is to
add ``walls'' that force the given $W\times H$-grid
to be divided into $k$ regions that have $B$ grid points available in each.
(These walls are simply high-degree vertices, but since every grid-point
must be used, nearly all neighbours of such a vertex must be in the layer 
below.)
Finally we add gadgets that encode $r_i,b_i,g_i$ in such a way that $b_i$
vertices must be in the $i$th region defined by the walls, while $r_i$ and
$g_i$ can freely choose into which of the regions they fall.  Therefore,
the division of them into the regions gives rise to a solution to
N3DM.

\paragraph{\bf Construction of $T$:}
Given an instance $\mathcal{I}$ of N3DM, we construct 
a tree $T$ with height $2k{+}3$ as follows. We start with the {\em spinal path} $v_1,\ldots,v_{2k+3}$ and choose $v_1$ to be the root of $T$.  
We add two {\em supporting paths} $u_2, \ldots, u_{2k+3}$ and $w_2, \ldots, w_{2k+3}$  where $u_2$ and $w_2$ are children of $v_1$.
We set $H:=2k+3$; hence in any strictly-upward drawing, $u_i,v_i,w_i$ must be 
on
 layer $\ell_i$ for $2\leq i\leq 2k+3$. 
(We count the layers from top to bottom,
i.e., layer $\ell_1$ is the layer with $y$-coordinate
$H$ that contains the root $v_1$,
and $\ell_i$ is the layer whose $y$-coordinate is $H{-}i+1$.)

\begin{figure}[pt]
\centering
\includegraphics[width=\textwidth]{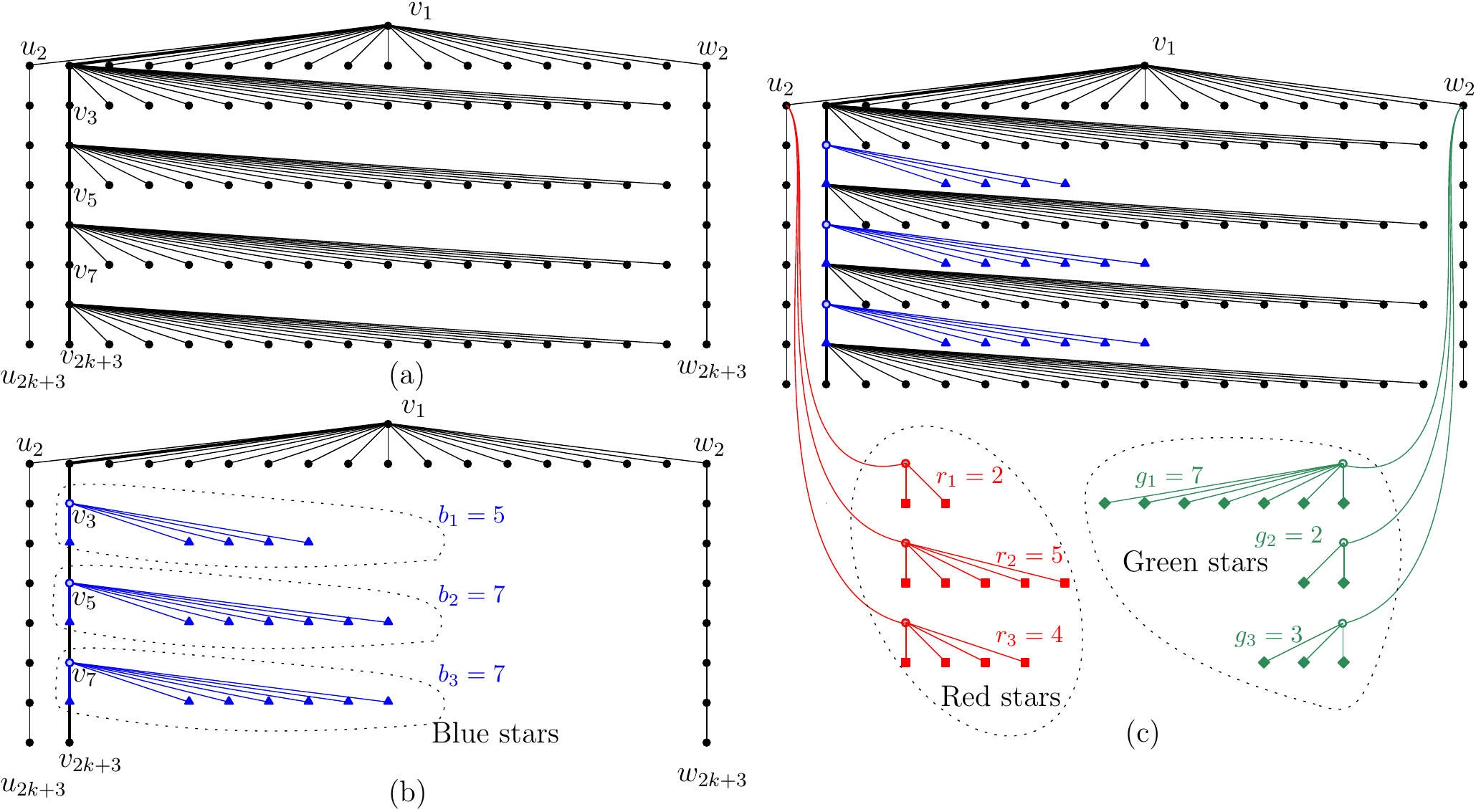}
\caption{(a) Illustration for the spinal path, supporting paths and walls.
 (b) Construction of the blue vertices. (c) Construction of the red and green vertices. 
 For space reasons, the numbers in this example do not satisfy the constraints that we imposed in N3DM.}
\label{fig:construction}
\end{figure}

Next we add the ``walls'' as shown in Fig.~\ref{fig:construction}(a). 
Namely, add $B+1$ leaves to $T$ that are children of the root; the star graph 
induced by $v_1$ and its children is called the \emph{wall of $v_1$}
with {\em wall root} $v_1$.  The {\em wall children} of $v_1$ are these
$B{+}1$ added leaves, as well as child $v_2$.
Similarly we add a {\em wall of $v_{2j}$}
for $j\in \{1,\dots,k+1\}$, by adding $B{+}1$ leaves to  
$T$ that are children of $v_{2j}$ so that $v_{2j}$ has $B{+}2$ wall children
(including $v_{2j+1}$).  
 
Finally, we encode
$r_i,g_i,b_i$ of  the N3DM instance. For $1\leq i\leq k$, we add $b_i{-}1$  
leaves  to $T$ and make them children of $v_{2i+1}$, see 
Fig.~\ref{fig:construction}(b).  We will call $v_{2i+1}$  a \emph{blue parent}  and its $b_i$ children (including $v_{2i+2}$) the \emph{blue children} of $v_{2i+1}$. For each $r_i$  (resp., $g_i$), we create a star with $r_i$ (resp., $g_i$)  leaves and connect the center to $u_2$ (resp., $w_2$), see Fig.~\ref{fig:construction}(c). We  refer to the stars corresponding to $r_i$ and $g_i$ as the \emph{red}  and \emph{green stars}, respectively. 
This finishes the construction of tree $T$.  Set $W:=B+4$ and
observe that $T$ has $2kB+2B+8k+9 = (B+4)(2k+2) + 1 = W\times (H-1)+1$ vertices,
which means that in any strictly-upward drawing in a $W\times H$-grid,
all layers except the top one must be completely full because the top layer 
can contain only one vertex (the root $v_1$).

\begin{figure}[pt]
\centering
\includegraphics[width=\textwidth]{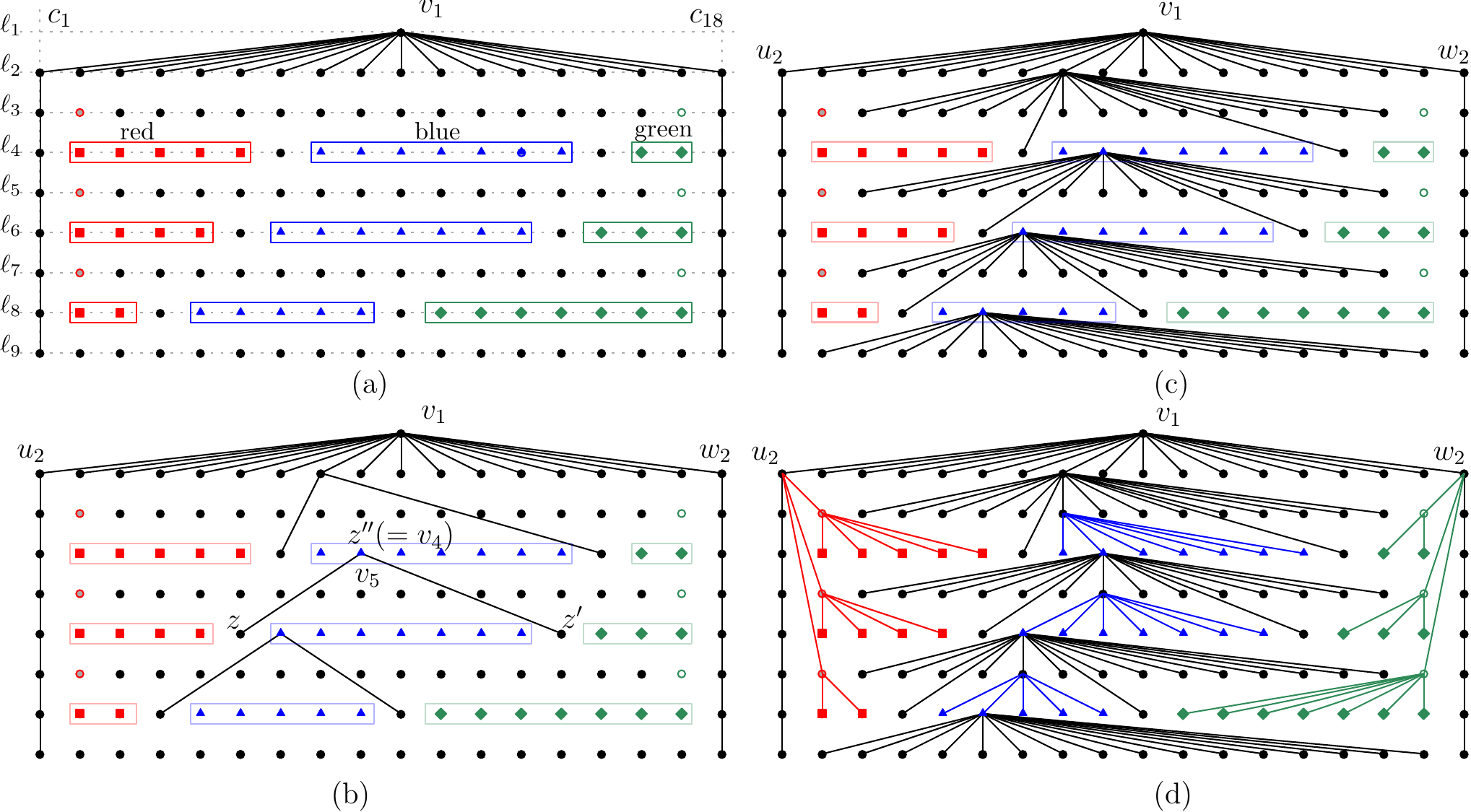}
\caption{ Construction of a drawing of $T$ on a $W\times H$ grid. 
 (a) Placement of the vertices of $T$,  drawing of the edges of the supporting paths, and the wall of $v_1$. (b)--(c) Drawing of the blue stars and the remaining walls. (d) Drawing of the red and green stars. 
}
\label{fig:easy}
\end{figure}

\paragraph{\bf From N3DM to tree drawing:}
If $\mathcal{I}$ is a yes-instance, then we create
a straight-line strictly-upward drawing of $T$
on  a $W\times H$ grid as illustrated in  Fig.~\ref{fig:easy}.
The root $v_1$ is anywhere in the top layer.
We place the supporting paths
in layers $\ell_2,\dots,\ell_{2k+3}$ in the leftmost (resp., rightmost) column,
as forced by the strictly-upward constraints.  Vertex $v_1$ has $B+4$
children ($u_2,w_2$ and the $B+2$ wall children); we place all
these in the second layer.    We do not yet pick which of these $B+2$ wall
children becomes $v_2$; this will be determined later.

The solution to $\mathcal{I}$ gives two
permutations $\pi,\pi'$.
Place the red, blue and green parents corresponding to $r_{\pi(i)}$,  $b_i$ and  $g_{\pi'(i)}$ on layer $\ell_{2i+1}$.  More precisely,
from left to right, we have first $u_{2i}$, then the red parent, 
then $B$ wall children of $v_{2i}$, then the green parent, and then $w_{2i}$. 
Later, we will choose one of these $B$ children to be $v_{2i+1}$, hence the blue parent corresponding to $b_i$.  Observe that layer $\ell_{2i+1}$ is completely filled with vertices of $T$.  

Place the red/blue/green children corresponding to $r_{\pi(i)}$, $b_i$ and $g_{\pi'(i)}$ on layer $\ell_{2i+2}$.     More precisely,
from left to right, we have first $u_{2i}$, then $r_{\pi(i)}$ red children, then one
wall child of $v_{2i}$, then $b_i$ blue children of $v_{2i+1}$, then
another wall child of $v_{2i}$, then $g_{\pi'(i)}$ green children and
finally $w_{2i}$.
Since $r_{\pi(i)}+b_i+g_{\pi'(i)} = B$,  layer $\ell_{2i+2}$ is also 
completely filled.   

We must argue that all edges can
be connected straight-line.  This holds for the red stars  since red
children lie in the layer below their parent, and red parents lie
in the column next to $u_2$.
See also Fig.~\ref{fig:easy}(d). 
Similarly we can connect green stars.  
For the blue stars, because the $b_i$'s are much larger than the $r_i$'s,
one can argue that the blue intervals of children of $v_{2i-1}$ and $v_{2i+1}$
overlap.  We pick $v_{2i}$ to be within this overlap in such a way
that it can connect it to the two wall children
that are on layer $\ell_{2i+2}$ without using up grid points. We use as 
$v_{2i+1}$ the point
 directly below $v_{2i}$; due to the choice of 
$v_{2i}$ then $v_{2i+1}$ can connect to
the blue interval on layer $\ell_{2i+2}$ without crossing.
 Details are 
\ifarxiv
in the appendix.
\else
in the full version~\cite{arxiv}.
\fi

\paragraph{\bf From tree drawing to N3DM:}

We now show that any strictly-upward polyline drawing of $T$ in a $W\times H$-grid gives rise to permutations $\pi$ and $\pi'$ such that  $r_{\pi(i)}+b_i+g_{\pi'(i)} = B$ holds for $1\leq i\leq k$.   We select $r_{\pi(i)}$ and $g_{\pi'(i)}$ in a bottom-up fashion, i.e., we first construct the triple $(r_{\pi(k)},b_k,g_{\pi'(k)})$, then the triple $(r_{\pi({k-1})},b_{k-1},g_{\pi'({k-1})})$, and so on.

Since $H$ equals the height of the tree, we know that
  $v_i,u_i,w_i$ (for $i>1$) are on layer $\ell_{i}$, and the wall children of
 $v_{2k+2}$ are on layer $\ell_{2k+3}$ (the bottommost layer).  Hence layer $\ell_{2k+3}$ contains
these $B+2$ wall children, as well as $u_{2k+3}$ and $w_{2k+3}$.
By $W=B+4$ this layer is full and contains no other vertices. 
 Also $v_{2k}$ lies is on layer $2k$, and so all its wall
 children must be on layers $\ell_{2k+1}$ and $\ell_{2k+2}$.
 We need an observation that crucially requires that all grid points are used.
 Details are 
\ifarxiv
in the appendix.
\else
in the full version~\cite{arxiv}.
\fi

\begin{lemma}\label{lem:abc}
Presume we know that all wall children of $v_{2i}$ are on layers $\ell_{2i+1}$
and $\ell_{2i+2}$.  	Then at most two wall children of $v_{2i}$ are on
layer $\ell_{2i+2}$, and the wall children of $v_{2i}$ on layer $\ell_{2i+1}$
occupy a consecutive set of points.
\end{lemma}

Thus there are at least $B$ wall children of $v_{2k}$ that form an interval 
on $\ell_{2k+1}$.  Also $u_{2k+1}$ and $w_{2k+1}$ are on layer $\ell_{2k+1}$,
leaving at most two points in this layer free to be used for 
 red or green stars, or blue or wall children from higher up.

We argue that indeed these two points must be used for a red parent and a
green parent. To see this, consider the interval $\lambda_m$ that consists of the middle 
$W{-}16{=}B{-}12$
\todo{Change of $\lambda_m$ here; we need it to be nearly everything!}
points on layer $\ell_{2k+2}$.   $T$ has $O(k^{2c+1})$ vertices
that are not in green stars while $B>3k^c+k^{2c}+k^{4c}$ 
so there must exist a green vertex in $\lambda_m$.  It must be a green
leaf $l_g$ since layer $\ell_{2k+3}$ is full. 

We claim that if a vertex $x$ uses a point in $\lambda_m$, then its parent
$p_x$ is either $v_{2k}$ or $p_x$ is on layer $\ell_{2k+1}$, i.e., one layer 
above $x$.
\todo{We had the same claim here twice, once for $l_g$ and once for $l_r$.  Replaced by doing it just once with an $x$.}
To see this, assume otherwise and consider the place where edge $(x,p_x)$
traverses layer $\ell_{2k+1}$.  If $p_x\neq v_{2k}$ then this point must be 
outside the interval of
 the $B=W-4$ wall children of $v_{2k}$ in this layer, 
else we would intersect 
 an edge.  So this point is within the leftmost or 
rightmost four units
 of $\ell_{2k+1}$.  Since $\lambda_m$ covers all but 
the leftmost and rightmost 8 points of layer $\ell_{2k+2}$, 
this forces $p_x$ to be outside the allotted width of the drawing, a 
contradiction. 
\ifarxiv
Fig.~\ref{fig:details}(b) in Appendix~\ref{app:choose} illustrates this scenario. 
\else
See the details in the full version~\cite{arxiv}.%
\fi

 Consequently, the green parent $p_g$ of $l_g$ is on layer $\ell_{2k+1}$
and its green children are {\em all} on $\ell_{2k+2}$. 
Due to our assumptions on N3DM, these green children, plus the blue children
of $v_{2k+1}$, are not enough to fill $\lambda_m$, but leave too little
space to place another green star. Therefore some point
in $\lambda_m$ must be occupied by a red child $l_r$, and its parent 
$p_r$
 hence is in layer $\ell_{2k+1}$. 
 Let $r_{\pi(k)}$ and $g_{\pi'(k)}$ be the numbers corresponding to the
red and green stars at $p_r$ and $p_g$.

We had $u_{2k+1},w_{2k+1}$,$p_r$,$p_g$,
and at least $B$ wall children  in $\ell_{2k+1}$, which by $W=B{+}4$ 
means that
there are exactly $B$ wall children in $\ell_{2k+1}$ and no other vertices.
So two wall children
are in layer $\ell_{2k+2}$.  These two, plus $u_{2k+2}$ and $w_{2k+2}$,
leave $B$ points for the red, blue and green children,
so $r_{\pi(k)}+b_k+g_{\pi'(k)}\leq B$.  On the other hand, $\lambda_m$ 
cannot contain any vertex $x$ other than
two wall-children of $v_{2k}$ and these red, blue and green children, because
layer $\ell_{2k+1}$ has no space left for the parent of $x$.
So $r_{\pi(k)}+b_k+g_{\pi'(k)}\geq B-14$. Since all input numbers are big
\todo{Here was a small hole: we never argued why these children {\em must}
fill the layer.  Fixed now.}
enough, this implies $r_{\pi(k)}+b_k+g_{\pi'(k)} = B$, as desired.

It also follows that $\ell_{2k+1}$ is completely filled by these vertices,
which means that no wall children of $v_{2k-2}$ can be in layer $\ell_{2k+1}$
or below. We can now apply the same argument iteratively to compute the upper level triples $(r_{\pi({k-1})},b_{k-1},$ $g_{\pi'({k-1})}), \ldots, (r_{\pi({1})},b_{1},g_{\pi'({1})})$. This completes the NP-hardness reduction. 

We can extract $\pi$ and $\pi'$ from any polyline drawing, while
a solution to $\mathcal{I}$ gives rise to a straight-line drawing.
Therefore, the reduction holds both in polyline and straight-line drawing settings. Since SUTD is clearly in NP we hence have:

\begin{theorem}
Given a tree $T$, and two natural numbers $W$ and $H$, it is NP-complete to decide whether $T$ admits a
 strictly-upward (polyline or straight-line) drawing with width  at most $W$ and height at most $H$. 
\end{theorem}

\section{Directions for Future Research}
Several interesting questions remain. 
 In our reduction, we crucially used that the underlying 
 tree has high degree, that the height of the drawing equals the
height of the tree,
that the order among children is not fixed,
and that the drawing is {\em strictly} upward. 
 Does SUTD remain NP-hard for bounded degree trees? 
 What if the given width is optimal for the tree and the height 
 needs to be minimized? 
How about order-preserving drawings and/or upward drawings?

The above questions are open also for many other popular styles 
  for drawing trees. Specifically, is the problem of computing
 (not necessarily upward) drawings of trees on a given 
 grid NP-hard? 
 Are there polynomial-time algorithms that can approximate the area
 within a constant factor? 

\bibliographystyle{splncs03}
\bibliography{bibs}
\ifarxiv
\newpage
\appendix

\section{Numerical 3-Dimensional Matching}
\label{app:N3DM}

We prove the NP-hardness of SUTD by a reduction from a strongly NP-complete
 problem numerical 3-dimensional matching~\cite{gANDj}, which is defined as follows.
\begin{enumerate}
\item[]\textbf{Problem:} \textsc{Numerical 3-Dimensional Matching} (N3DM)
\item[]\textbf{Instance:} Positive integers $r_i,b_i,g_i$, where $1\le i\le k$, and an integer $B$ such that 
  $\sum_{i=1}^k (r_i+b_i+g_i) = k\cdot B$.  
\item[]\textbf{Question:} Do there exist permutations $\pi$ and $\pi'$  such that
  for each $i$, the equality $r_{\pi(i)}+b_i+g_{\pi'(i)} = B$ holds?
\end{enumerate}

We now show that N3DM remains NP-complete even under the following constraints: 
 $b_i$'s are odd, $3k^c \le r_i \le 4k^c$, $k^{2c} \le b_i \le k^{2c}+k^c$, 
 and $k^{4c} \le g_i\le k^{4c}+k^c$, where $c>1$ is a constant.  
 
Given a N3DM instance $\mathcal{I}$, we construct an equivalent N3DM instance $\mathcal{I}'$
 that satisfies the above constraints. 
 Since N3DM is strongly NP-complete, we can choose some constant $c$ such that 
 all the numbers of $\mathcal{I}$ are bounded by 
 $(k^c-2)/2$. For each $r_i,b_ig_i$, we construct corresponding the numbers 

\begin{align}
 r'_i &= 3k^c+2r_i,\\
 b'_i &= k^{2c}+2b_i+\mu,\\
 g'_i &= k^{4c}+2g_i,
\end{align}
 where $\mu$ is 0 or 1 depending on whether $k$ is odd or even.
 Note that $b'_i = (k^{2c}+2b_i+\mu)$ is odd irrespective of the parity of $k$.
 Since $r_i,b_i,g_i$ are bounded by $(k^c-2)/2$, 
 all the required constraints are now satisfied. 
 
Choose $B'$ to be $(3k^c + k^{2c} + k^{4c} + 2B+1)$. For any 
 $r_{\pi(i)}+b_i+g_{\pi'(i)} = B$, the corresponding numbers in $\mathcal{I}'$
 sum to 
\begin{align}
&3k^c + 2r_{\pi(i)}+ k^{2c}+2b_i+1 k^{4c}+2g_{\pi'(i)}\\
&= 3k^c + k^{2c} + k^{4c} + 2B+1,\\
&= B'.
\end{align}
 
On the other hand, any triple that sum to $B'$ in $\mathcal{I}'$ 
  corresponds to a triple in $\mathcal{I}$ that sum to $B$.

\section{Choosing $v_{2i}$ among blue children}
 \label{app:choose}
It remains to say how
we choose blue parents so that we can connect the blue stars.
Note that for $i>1$ layer $\ell_{2i}$ contains an interval of $b_i$ 
blue children of $v_{2i-1}$.   We can expand this notation to $i=1$ if
we consider the interval
formed by the $B+2$ wall children of $v_1$ to be a blue interval as well,
so now consider $i\geq 1$.
We must choose $v_{2i}$ among the blue interval in layer $\ell_{2i}$
so that it can connect
to the two wall children that were placed two layers below 
without using up a grid point, and such that there is a place for $v_{2i+1}$
where it can connect both to $v_{2i}$ and to its blue children.

Observe that all the red children corresponding to $r_i$ 
 appears consecutively before the blue children.
 Since $3k^c \le r_i \le 4k^c$ and $k^{2c} \le b_i \le k^{2c}+k^c$,
 the blue intervals on layer $\ell_{2i}$ and $\ell_{2i+2}$
 start at $x$-coordinate $4k^c+3$ or farther left and end at
	$x$-coordinate $3k^c+k^2c+2$ or farther right, and hence
 overlap in an interval $\lambda$ that has length at least $k^{2c}-k^c\geq 2$
(we assume here and elsewhere that $k$ is sufficiently big).
 See Fig.~\ref{fig:details}(a). 
 We claim that either the leftmost or second left
point of $\lambda$ can be used for $v_{2i}$ and argue this as follows.

\begin{figure}[h]
\centering
\includegraphics[width=\textwidth]{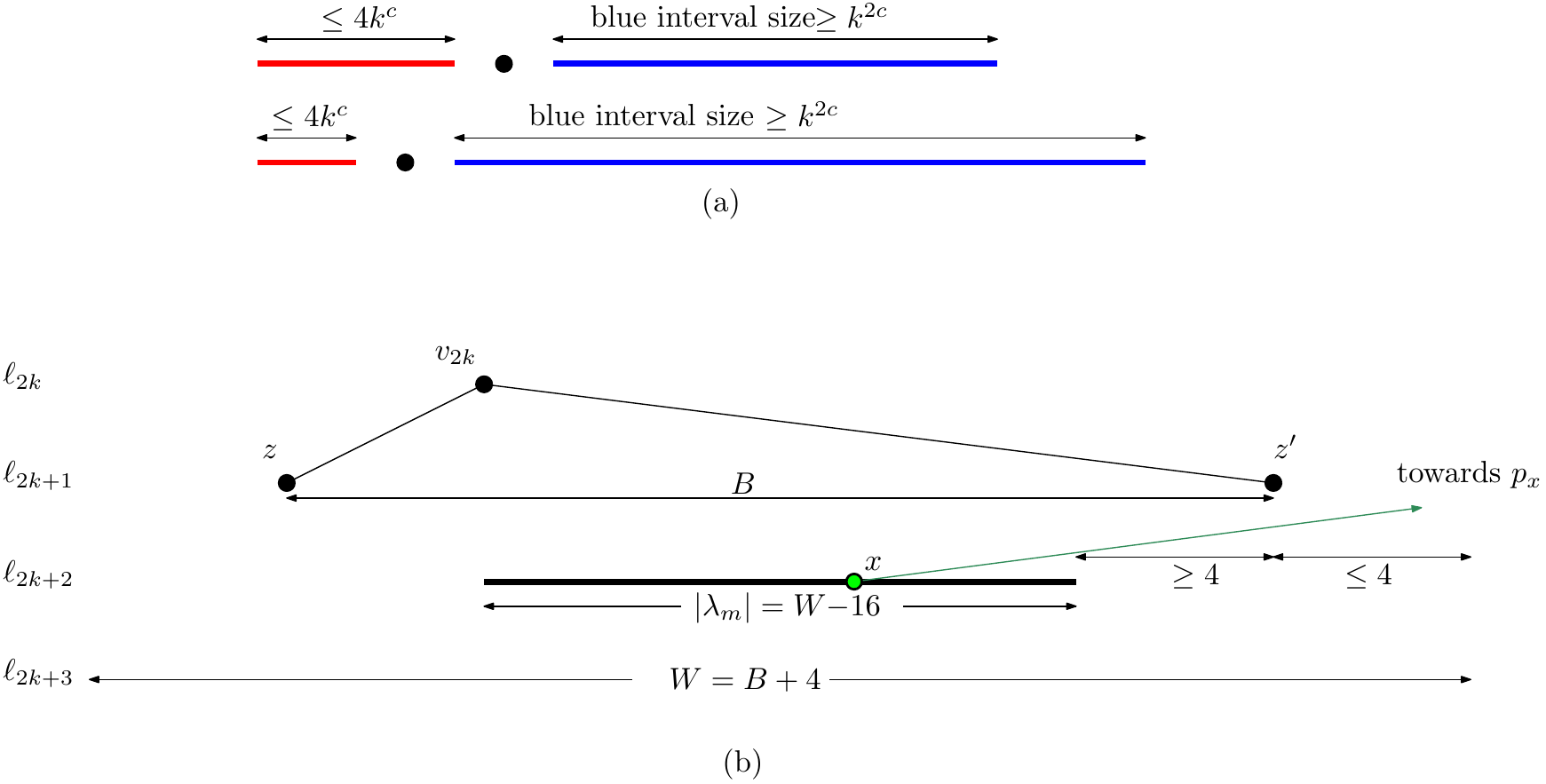}
\caption{(a) Intersection of blue intervals. (b) Placement of  $p_x$.}
\label{fig:details}
\end{figure}

Let $z,z'$ be the two wall children of $v_{2i}$ that were placed on
layer $\ell_{2i+2}$; they are both outside the range $\lambda$ since they
have $b_i$ points between them.
Since $b_i$ is odd, the horizontal distance $\delta_x(z,z')$ between
$z$ and $z'$ is even.  So the first or second grid point $z''$ of $\lambda$
satisfies that $\delta_x(z,z'')$ and  $\delta_x(z',z'')$ are  odd. 
Choose $z''$ to be $v_{2i}$ and the grid point immediately below $z''$ as 
$v_{2i+1}$, e.g., see Fig.~\ref{fig:easy}(b). Since $v_{2i+1}$
falls within range $\lambda$,
$v_{2i+1}$ sits between the edges from $v_{2i}$
to $z,z'$ and therefore can connect to its blue children without crossing.
 Consequently, we obtain the desired drawing of $T$.

\section{$T$ uses all points}

This section gives the detailed computation why $T$ must use all grid
points that it can use.  Recall that the spinal path has $2k+3$ vertices
while the supporting paths 
contain $2\cdot(2k+2) = (4k+4)$ vertices. There are $(k+1)$ wall roots on the spinal path $P_s$, and each is adjacent to $(B+2)$ wall children. 
Finally, $T$ has   $2k+\sum_{i=1}^k (r_i+b_i+g_i) = 2k+k\cdot B$ vertices corresponding to the numbers of $\mathcal{I}$, where $2k$ corresponds to the red and green roots. Note that for each blue root, one of its blue children is a wall root, which is double counted. Therefore, the total number of vertices in $T$ is 
 $$(4k+4) + (2k+3)+(k+2)(B+1)+ 2k + k\cdot B - k
  = 2kB+2B+8k+9 = (B+4)(2k+2) + 1,$$
which is exactly $W\times (H-1)+1.$
%

\section{Proof of Lemma~\ref{lem:abc}}
\label{app1}

\noindent\textbf{Lemma~\ref{lem:abc}}\emph{
Presume we know that all wall children of $v_{2i}$ are on layers $\ell_{2i+1}$
and $\ell_{2i+2}$.  	Then at most two wall children of $v_{2i}$ are on
layer $\ell_{2i+2}$, and the wall children of $v_{2i}$ on layer $\ell_{2i+1}$
occupy a consecutive set of points.
}
\begin{proof}
  By construction, every wall root has $(B+2)$ wall children. 
 We first show that the wall children placed on layer  $\ell_{2i+1}$ must be
 consecutive.

If the  wall children placed on layer  $\ell_{2i+1}$ are not placed consecutively, then there are
 three grid points $q_l,q,q_r$ on $\ell_{2i+1}$  in this order from left to right such that each
 of  $q_l,q_r$ contains a  wall child, but $q$ does not contain any wall child. 
Since all points must be used,  $q$
 must contain a vertex $a$ of $T$. Since the parent  of $a$ must lie on $\ell_{2i}$
 or above, it will intersect either the edge $(v_{2i},q_l)$ or the edge $(v_{2i},q_r)$, a 
 contradiction.

Now we show that there must be at least $B$ wall children of $v_{2i}$
on $\ell_{2i+1}$. 
Suppose for a contradiction that $\ell_{2i+1}$ contains fewer than $B$ of them, and hence $\ell_{2i+2}$ contains three or more wall children of $v_{2i}$. 
Let $z_a,z_b,z_c$ be three grid points containing wall children of $v_{2i}$ on $\ell_{2i+2}$, in this order from left to right, and chosen such that no 
 point inbetween them contains wall children of $v_{2i}$.

The edges connecting $v_{2i}$ to $z_a,z_b,z_c$ cannot have bends, for any bends would use up a grid-point that is needed for a vertex of $T$.
Furthermore, the horizontal distance $\delta_x(v_{i},z)$ is odd for all 
$z\in \{z_a, z_b,z_c\}$.  Otherwise, if 
$\delta_x(v_{i},z)$ were even then the straight line $v_{2i}z$ would 
traverse $\ell_{2i+1}$ at an integer $x$-coordinate, and there use up a
grid-point that we need for the vertices of $T$, as 
 illustrated in gray circle in Fig.~\ref{fig:lem1}.
\begin{figure}[pt]
\centering
\includegraphics[width=\textwidth]{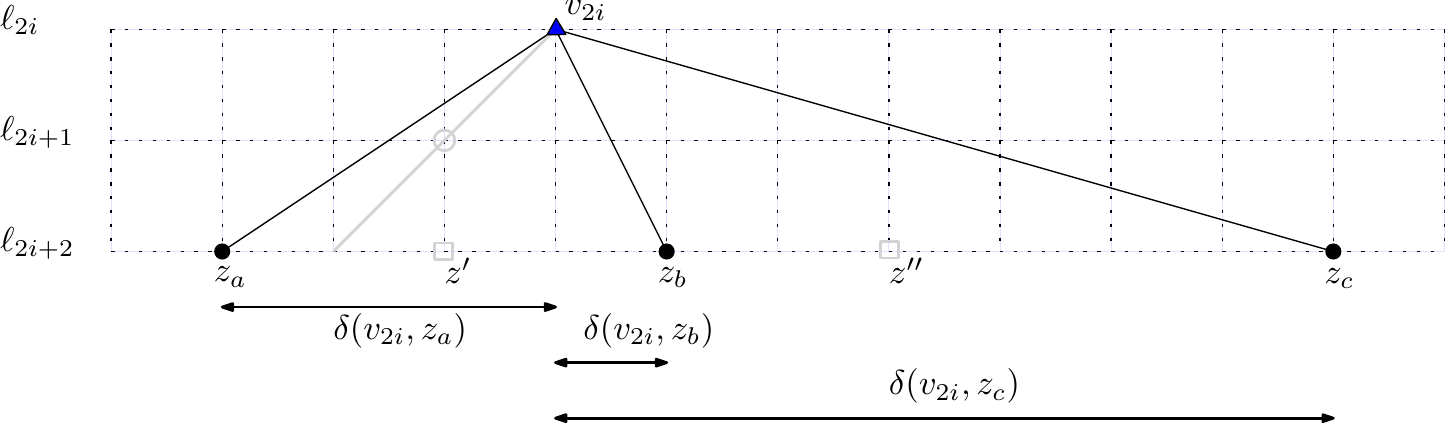}
\caption{(a) Illustration for Lemma~\ref{lem:abc}.  }
\label{fig:lem1}
\end{figure}

Since  $\delta_x(v_{i},z)$ is odd for $x\in \{z_a,z_b,z_c\}$, we can find a grid point $z'$  between $z_a,z_b$ and a grid point $z''$ between $z_b,z_c$, e.g., see Fig.~\ref{fig:lem1}.  Some vertices $q',q''$ of $T$ reside on 
$z'$ and $z''$ respectively, and they are not wall children of $v_{2i}$
by choice of $z_a,z_b,z_c$. Since $\Gamma$ is a planar drawing, the path from root to $q'$ (similarly, from root to $q''$)  must pass through $v_{2i}$. Therefore, $v_{2i}$ must have two non-leaf children, a contradiction.
\end{proof}

\fi 
\end{document}